%% file: main.tex
\documentclass{article}
\usepackage{spconf,amsmath,graphicx}

\usepackage{enumitem}
\setlist{nosep, leftmargin=14pt}

\usepackage{multirow}
\usepackage{bbding}
\usepackage{pifont}
\usepackage{wasysym}
\usepackage{amssymb}
\usepackage{tabularx,ragged2e}
\newcolumntype{C}{>{\arraybackslash}X}
\usepackage{graphicx}
\usepackage{rotating}
\usepackage{graphicx}
\usepackage{booktabs}
\usepackage{multirow}
\usepackage{rotating}
\usepackage{lscape}
\usepackage{booktabs}
\usepackage{xcolor,colortbl}
\usepackage{makecell}

\usepackage{lipsum}% http://ctan.org/pkg/lipsum
\usepackage{algpseudocode}% http://ctan.org/pkg/algorithmicx
\usepackage{amsmath}
\usepackage[linesnumbered,ruled]{algorithm2e}
\definecolor{lavender}{gray}{0.9}
\usepackage{array}
\usepackage{mwe} % to get dummy images
\usepackage{graphicx}
\usepackage{todonotes}
\usepackage{bbding}
\usepackage{booktabs}
\usepackage{multirow,bigdelim}
\usepackage{color}
\usepackage{xr}

\usepackage{amsmath,amsfonts,amssymb}

\usepackage{rotating}
\usepackage{mwe}
\usepackage{graphbox}
\usepackage{multirow}
\usepackage{amsfonts} 
\usepackage{amsmath,mathtools}
\usepackage{amssymb}
\usepackage{bm}
\usepackage{url}
\usepackage[colorlinks=true, allcolors=black]{hyperref}
\def\arrvline{\hfil\kern\arraycolsep\vline\kern-\arraycolsep\hfilneg}

\title{Promise: \underline{Pro}mpt-Driven  3D \underline{M}edical \underline{I}mage \underline{Se}gmentation Using Pretrained Image Foundation Models}
\name{Hao Li, Han Liu, Dewei Hu, Jiacheng Wang, Ipek Oguz}
\address{Vanderbilt University} 
\begin{document}
%\ninept
%

\maketitle
\begin{abstract}
To address prevalent issues in medical imaging, such as data acquisition challenges and label availability, transfer learning from natural to medical image domains serves as a viable strategy to produce reliable segmentation results. However, several existing barriers between domains need to be broken down, including addressing contrast discrepancies, managing anatomical variability, and adapting 2D pretrained models for 3D segmentation tasks.
% However, several existing barriers between domains need to be broken down: (1) notable discrepancy in contrast, (2) complex anatomical variability in medical images, and (3) 2D pretrained image foundation models may struggle with 3D medical segmentation task.
In this paper, we propose ProMISe, a prompt-driven 3D medical image segmentation model using only a single point prompt to leverage knowledge from a pretrained 2D image foundation model. 
% In this paper, we propose Promise, a promptable model, for 3D medical image segmentation and utilize the knowledege from pretrained 2D image foundation model.
In particular, we use the pretrained vision transformer from the Segment Anything Model (SAM) and integrate lightweight adapters to extract depth-related (3D) spatial context without updating the pretrained weights. 
% Furthermore, we employ shallow networks for prompt generator and mask decoder tailored for volumetric medical segmentation. 
For robust results, a hybrid network with complementary encoders is designed, and a boundary-aware loss is proposed to achieve precise boundaries. We evaluate our model on two public datasets for colon and pancreas tumor segmentations, respectively. Compared to the state-of-the-art segmentation methods with and without prompt engineering, our proposed method achieves superior performance.
The code is  publicly available at \url{https://github.com/MedICL-VU/ProMISe}
\end{abstract}

% \begin{keywords}
% Medical image segmentation, 3D adapter, transfer learning, pretrained image foundation model
% \end{keywords}

\section{Introduction}
\input{text/_1-introduction}

% \section{Related works}
% \input{text/-2-related_works}

\section{Methods}
\input{text/_3-methods}

\section{Experiments}

\input{text/_4-experiments}

\section{conclusion}

\input{text/_5-conclusion}

\clearpage
\section{Compliance with Ethical Standards}
This research study was conducted retrospectively using human subject data made available in open access by MSD. Ethical approval was not required as confirmed by the license attached with the open access data.

\bibliographystyle{IEEEbib}
\bibliography{refs}

\end{document}

%% file: text/_1-introduction.tex
Recently, image segmentation foundation models \cite{kirillov2023segment,zou2023segment} have revolutionized the field of image segmentation, demonstrating wide generalizability and impressive performance by training on massive amounts of data to learn general representations. 
% For instance, segment anything model (SAM) \cite{kirillov2023segment} trained with 1.1 billion segmentation masks collected from 11 million natural images. 
% For instance, segment anything model (SAM) \cite{kirillov2023segment} trained from 11 million natural images. 
Prompt engineering further improves the segmentation capability of these models. Given proper prompts as additional inputs, these models can handle various zero-shot tasks across domains and produce reliable segmentations during inference. Unlike these broad successes, medical image segmentation is often limited by  issues such as expensive data acquisition and time-consuming annotation processing, resulting in a lack of massive public datasets available for training. Thus it is desirable to leverage transfer learning from the natural image domain for robust medical image segmentation \cite{matsoukas2022makes}.

However, directly leveraging pretrained 2D natural image foundation models for 3D medical image segmentation often leads to sub-optimal results \cite{he2023computervision}. This is primarily because: (1) medical images have their own unique contrast and texture characteristics; (2) anatomical differences among individuals make medical image segmentation challenging; and (3) slice-wise (2D) segmentation with transfer learning discards important depth-related spatial context in 3D medical data. Given these challenges, can we effectively adapt the pretrained models to achieve robust 3D medical segmentations?

\begin{figure*}[t]
\centering
\includegraphics[width=1\linewidth]{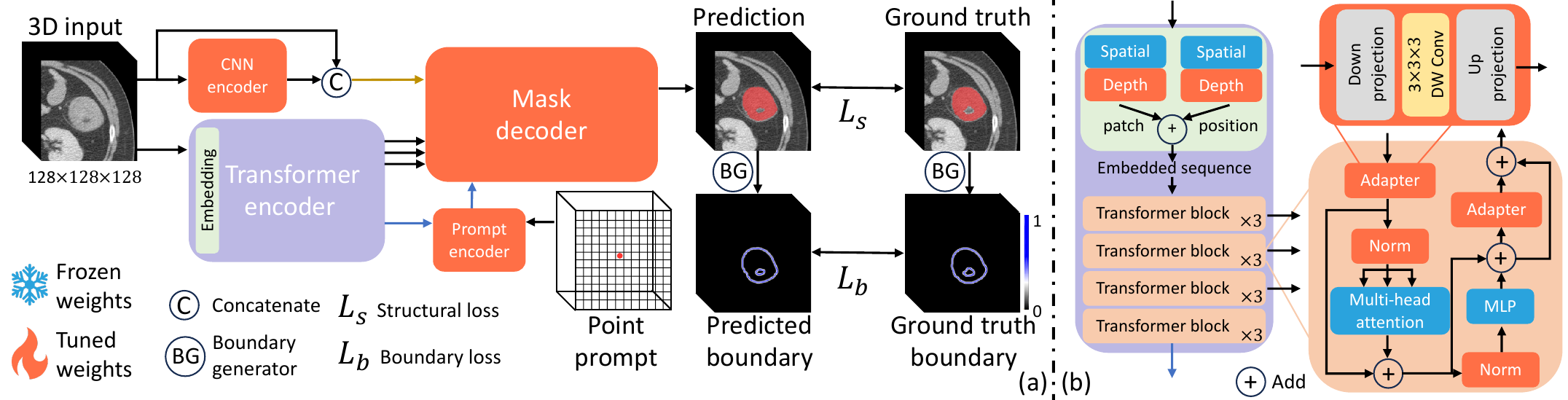}   
\caption{The proposed framework (ProMISe) and details of transformer encoder are shown as (a) and (b), respectively.}
\label{framework}
\end{figure*}

% In this paper, we propose ProMISe, a \textbf{pro}mpt-driven hybrid network for 3D \textbf{m}edical \textbf{i}mage \textbf{se}gmentation by adapting 2D pretrained image foundation models
In this paper, we propose ProMISe, \textbf{pro}mpt-driven  3D \textbf{m}edical \textbf{i}mage \textbf{se}gmentation using pretrained image foundation models (see Fig.~\ref{framework}). Specifically, ProMISe takes a 3D input image and a single point prompt as inputs, and uses image and prompt encoders to produce segmentation. Unlike most promptable models, a shallow convolutional neural network (CNN) is used as complementary path alongside the pretrained transformer image encoder \cite{kirillov2023segment}, with adapters employed within the transformer to capture 3D depth context. During training, most weights of the adapted transformer encoder remain static; the other components in the proposed method are designed in a lightweight manner for efficiency and trained from scratch. We use a structural loss and a novel boundary-aware loss for precise decisions. %These aim to tackle challenging medical segmentation tasks. 
The main novel contributions are:
% (1) We propose a method for 3D medical image segmentation that adapts pretrained models. We employ plug-and-play lightweight adapters to better optimize knowledge transfer across domains and more effectively fine-grained features. Our approach is compatible with various pretrained image models, straightforward to implement, and cost-effective to train. 

% (2) We devise the model as a hybrid network, incorporating an additional CNN encoder and exploring diverse 3D network designs. Additionally, we proposed a boundary-aware loss alongside the segmentation loss. Every element is crafted to ensure robust segmentation. and (3) We evaluate the proposed Promise on two public datasets %\cite{antonelli2022medical} 
%   for colon and pancreas tumor segmentation. Compared to state-of-the-art (SOTA) segmentation methods, our proposed method has superior performance.
  % in terms of higher Dice and normalized surface Dice scores.

\begin{itemize}
  \item We propose a method for 3D medical segmentation that adapts pretrained image foundation models. Plug-and-play lightweight adapters are used to better optimize knowledge transfer across domains and more effectively capture fine-grained features. Our method is compatible with various pretrained image models, easy to implement, and cost-effective to train. 

  \item We present a simple yet efficient boundary-aware loss for ambiguous edges. This ready-to-use loss can be seamlessly integrated into any training process without the need for offline edge map generation from ground truth.

  \item We validate the performance on two public datasets for challenging tumor segmentations. Our method outperforms state-of-the-art segmentation methods consistently.
\end{itemize}

% \paragraph{\textbf{Related works.}}
% \textbf{Adapting image foundation models.}
\noindent\textbf{Related works.} Fully fine-tuning image foundation models for a task requires a large amount computational resources and is not training-efficient. In contrast, partially fine-tuning \cite{ma2023segment} or introducing and training new shallow layers, such as lightweight adapters \cite{pan2022st,chen2022adaptformer,gong20233dsam,yang2023aim,wu2023medical} and the Low-Rank Adaptation (LoRA) module \cite{hu2021lora,zhang2023customized}, have demonstrated robust performance as parameter-efficient fine-tuning methods.
Recent works use SAM \cite{kirillov2023segment} for 3D medical image segmentation in a 2D slice-wise manner, which discard important depth-wise (3D) information and may require additional efforts to create prompts \cite{ma2023segment,zhang2023customized}. Other models use adapters; this approach has proven effective for adapting a pretrained model from 2D images to 3D (2D+time) videos \cite{pan2022st,yang2023aim}, and it has subsequently been utilized in 3D medical image segmentation \cite{wu2023medical} with the use of adapters in the pretrained transformer block \cite{chen2022adaptformer}. Although these models can segment 3D medical images, the image encoder still operates in a slice-wise (2D) manner with an additional branch for depth information. The weights for this branch that are replicated from the spatial branch demand more computational resources. In contrast, a holistic adaptation of SAM for 3D medical segmentation was proposed in \cite{gong20233dsam}, which avoids a depth branch by including an adaptor with depth-wise convolution \cite{pan2022st}. However, a single adapter in each transformer block may not fully achieve accurate adaptation due to the notable discrepancies between natural and medical images. Moreover, this method struggles to adequately capture details  and can lead to sub-optimal results, especially for tumor segmentation. These challenges and the critical importance of precise segmentation in medical applications motivate our proposed model as a more robust solution.% for 3D medical segmentation. 

%% file: text/_3-methods.tex
%\noindent\textbf{Framework overview.}
Fig.~\ref{framework} illustrates ProMISe, our proposed framework for 3D medical image segmentation, which employs prompt engineering and a pretrained image foundation model. Specifically, a 3D patch is taken as input and is fed through complementary CNN and transformer encoders. The prompt encoder utilizes the deepest feature from the transformer encoder (blue arrow in Fig.~\ref{framework})
as input together with the point prompt. Subsequently, all features, including the original input, are used to predict the segmentation mask via a lightweight CNN decoder. During training, the transformer encoder is partially tuned, while the rest are trained from scratch.
% while the CNN encoder, prompt encoder, and mask decoder receive full tuning.

\noindent\textbf{Image encoders.} Our model is designed to effectively capture both global and local information using complementary transformer and CNN encoders, respectively. 

For the transformer encoder (Fig.~\ref{framework}(b)), the input 3D image patch first passes through an embedding layer to create tokens with their positional information. Specifically, the pretrained weights from SAM \cite{kirillov2023segment} are employed for spatial patch embedding, and we introduced a trainable depth embedding layer for 3D data. The same approach is applied for positional encoding. Furthermore, we adapted the pretrained weights from SAM and fine-tuned the normalization layer in every transformer block. Unlike other works that employ a single adapter at the beginning of the transformer block \cite{pan2022st,gong20233dsam}, 
an additional lightweight adapter is used before the output to optimize knowledge transfer across domains and further refine the image features. Notably, the adapter employs depth-wise convolution to handle 3D images.
% an additional lightweight adapter is used before the output to further refine the image features. It is worth noting that the adapter employs depth-wise convolution to accommodate 3D images. 

\begin{figure}[t]
\centering
\includegraphics[width=1\linewidth]{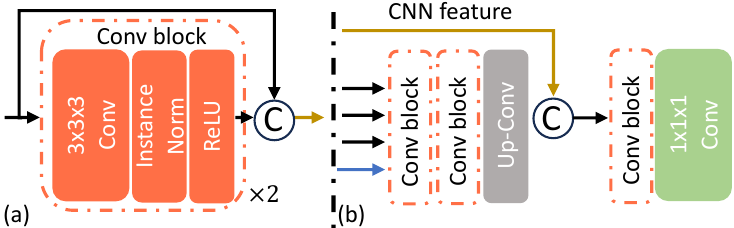}  
\caption{The details of (a) CNN encoder, and (b) mask decoder. 
% CNN and transformer highest features are denoted in yellow and blue, respectively.
}
\label{networks}
\end{figure}

Inspired by the hybrid network design \cite{li2022cats}, a CNN encoder is used to capture detailed information to complement the transformer. This is particularly desirable for tumor segmentation, as the boundaries are often ambiguous. It is designed as a shallow network for efficiency (Fig.~\ref{networks}(a)).

\begin{figure}[b]
\centering
\includegraphics[width=1\linewidth]{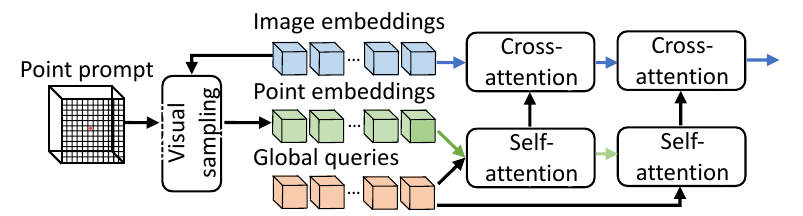}   
\caption{The details of the proposed prompt encoder.
% The output feature is mark in blue and fed to mask decoder.
}
\label{promptencoder}
\end{figure}

\noindent\textbf{Prompt encoder.}
We adapt the visual prompt encoder based on \cite{zou2023segment} (Fig.~\ref{promptencoder}). Unlike the prompt encoder proposed in SAM \cite{kirillov2023segment}, we incorporate image embeddings from the transformer encoder as an additional input. Point embeddings are derived from the given point prompt and image embedding using visual sampling (e.g., grid sampling) to ensure that their semantic features are aligned with image embeddings. Subsequently, the self-attention layer is applied to the point embeddings and learnable global queries. Afterwards, the image embeddings are applied to these queries via cross-attention. The output of the prompt encoder is fed to the mask decoder.% to generate final segmentation. 

During training, 10 random points from background are provided for each input patch to increase the generlizability to noisy prompts. In contrast to previous work that utilized 40 points from target region as prompts \cite{gong20233dsam}, we randomly select 10 point prompts during each iteration if the input patch contains foreground. For prompt engineering, our goal is a single click with minimal prior knowledge, but more prompts are supported if desired during inference.

\noindent\textbf{Mask decoder.} Instead of directly adapting the mask decoder from the foundation model in a 2D manner, we designed a shallow network to efficiently capture features in 3D and trained it from scratch (Fig.~\ref{networks}(b)). The multi-level features from the transformer encoder (Fig.~\ref{framework}(b)) are refined by two successive convolutional blocks. These are followed by a transposed convolution to ensure the features remain the same size. The fused features are processed through another convolutional block and a segmentation head for final results.

% Before producing the segmentation, the fused features are processed through another convolutional block and a segmentation head.

% \begin{figure}[t]
% \centering
% \includegraphics[width=0.9\linewidth]{image/network.pdf}   
% \caption{The details of proposed networks. \xxx{do all the arrow colors mean something? }}
% \label{networks}
% \end{figure}

\noindent\textbf{Boundary-aware loss.} In medical image segmentation, accurately delineating the boundaries of objects is important, especially for irregularly shaped objects such as tumors \cite{liu2023medical}. Besides popular structural segmentation losses, such as the combined Dice loss and cross-entropy loss (denoted as $L_{structural}$), we further propose a boundary loss ($L_{boundary}$) to preserve fine details and produce robust segmentations. Moreover, by emphasizing edge accuracy, the model might generalize better to unseen data for tumor segmentation. As shown in Fig.~\ref{framework}, we extract a smooth boundary map rather than a binary boundary for a more robust representation, and because learning from a binary boundary is a challenging task. Specifically, we use average-pooling operation $P_{ave}$ with kernel size 5 as boundary generator. Given a binary mask $M$, the smooth boundary is derived: $B(M) = |M - P_{ave}(M)|$. The total objective function is:
\begin{equation*}
\resizebox{\hsize}{!}{$
    L(S, G)= \lambda_1 L_{structural}(S, G) + \lambda_2 L_{boundary}(B(S), B(G))
    $}
\end{equation*}
    % \resizebox{\hsize}{!}
    %  {$L(S, G)= \lambda_1 L_{structural}(S, G) + \lambda_2 L_{boundary}(B(S), B(G))$}
where $S$ and $G$ represent segmentation and ground truth. $L_{structural}=L_{Dice} + L_{CE}$ is used to capture the structural information and $L_{boundary}=L_{MSE}$ recovers the detailed contours. Unlike other methods \cite{kervadec2019boundary} that require complicated offline computation of edge or distance maps to avoid iterative generation, our proposed ready-to-use boundary loss is computationally efficient and can be easily adapted to any segmentation task, and is independent of any augmentation.

%% file: text/_4-experiments.tex
\input{table/main_table}

\begin{figure*}[t]
\centering
\includegraphics[width=0.93\linewidth]{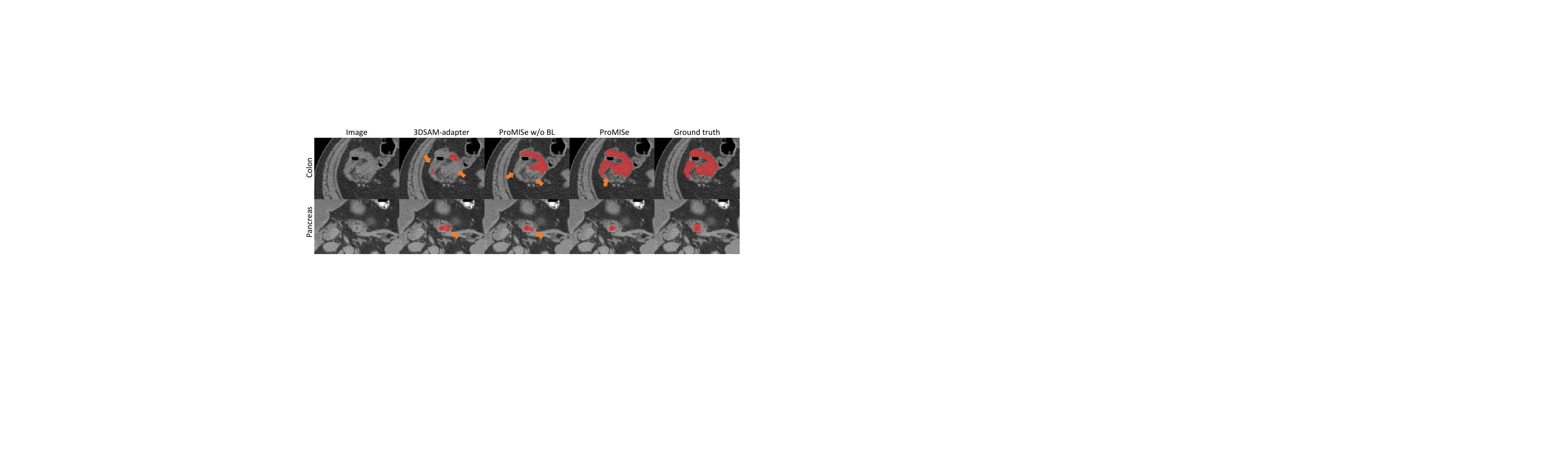}
\caption{Qualitative results. BL denotes boundary-aware loss. The major differences are highlighted by orange arrows.}
\label{qualitative results}
\end{figure*}

\input{table/ablation_table}

\subsection{Experimental settings}
\noindent\textbf{Datasets.}
We evaluated our proposed method on two public datasets from the Medical Segmentation Decathlon (\url{http://medicaldecathlon.com/}) for challenging tumor tasks from pancreas and colon applications, where ambiguous edges are present. These consist of 281 ($0.61\times0.61\times0.7$ to $0.98\times0.98\times7.5 mm^3$) and 126 ($0.54\times0.54\times1.25$ to $0.98\times0.98\times7.5 mm^3$) 3D CT volumes, respectively. Following the setup from the prior study \cite{gong20233dsam}, we used the same data split for each task with a training/validation/testing split of 0.7/0.1/0.2 and only use tumor labels to focus on binary segmentation.

\noindent\textbf{Preprocessing.}
% The preprossing is adapted from nnU-Net \cite{isensee2021nnu} pipeline. Briefly, this includes resampling, foreground $\pm0.5$ 
We resample to 1$mm$ isotropic resolution, intensity clip based on foreground $0.5$ and $99.5$
percentiles, and Z-score normalize based on all foreground voxels. Four data augmentations were used: random flip, rotation, zoom, and intensity shift. During training, an input patch of size $128\times128\times128$ was randomly selected such that its center pixel is equally likely to be foreground or background. Subsequently, each dimension was upsampled to 512.
% to optimally utilize the pretrained patch embedding layer.

\noindent\textbf{Implementation details.}
We utilized pretrained ViT-B from SAM \cite{kirillov2023segment} as transformer encoder, and set $\lambda_1 : \lambda_2 = 1: 10$ during training. The batch size was 1, and initial learning rate was 0.0004 with decreased amount $2{e^-6}$ every epoch. The AdamW optimizer was used with a maximum of 200 epochs. We used PyTorch, MONAI and an NVIDIA A6000 GPU for our experiments. The Dice score and normalized surface Dice (NSD) are used for evaluation. Compared state-of-the-art methods include: CNN (nnU-Net \cite{isensee2021nnu}), CNN with large kernel (3D UX-Net \cite{lee2023d}), Swin-encoder with CNN decoder (Swin-unetr \cite{tang2022self}), pure transformer (nnFormer \cite{zhou2023nnformer}), and adaptation method with adapters (3DSAM-adapter \cite{gong20233dsam}). We retrained using their official codes, and the pretrained weights are also employed if publicly available.

\subsection{Results} 

\noindent\textbf{Quantitative results.} Tab.~\ref{main table} presents a detailed comparison of results for colon and pancreas tumor segmentation. Notably, while CNN-based networks segment these tumors more effectively than transformers, prompt-driven methods outperform others when provided with only a single point in the entire volume. Our proposed method consistently outperforms all in terms of both Dice and boundary (NSD) metrics.

\noindent\textbf{Ablation study.} We also investigated the efficiency variations of the proposed ProMISe (Tab.~\ref{ablation study}). The use of two adapters and the boundary-aware loss mostly improved the results. Interestingly, switching from trilinear upsampling to up-convolution improved the performance for the colon, but showed a decline for the pancreas. This implies that trilinear upsampling may be more appropriate for pancreas tumors, which are typically round in shape. Using concatenation (-C) in the CNN encoder offers better Dice scores than residual connections (-R), though the latter improves surface quality more. While the performance of ProMISe improves with 10 prompts, the improvement is limited over a single prompt. Furthermore, it is challenging to identify the tumor area due to ambiguous boundaries, making the use of a single click preferable in practice, as it requires less expert knowledge.

\noindent\textbf{Qualitative results.} 
Fig.~\ref{qualitative results} shows qualitative visualizations from top-performing promptable methods. ProMISe yields results that closely align with the ground truth. 3DSAM-adapter \cite{gong20233dsam} fails to detect certain regions that ProMISe captures, even without the boundary-aware loss. This indicates the improved generalizability of the model through our proposed modifications. 
Moreover, the use of the boundary-aware loss yields robust segmentations, alleviating issues of both under-segmentation for colon and over-segmentation for pancreas tumors, respectively. 
Notably, the boundary-aware loss improves segmentation not just for the irregularly shaped colon tumors but also for the pancreas tumors, which typically have a more regular, rounded shape. However, slight under-segmented areas are found in pancreas segmentation.

% Despite these improvements, minor defects such as slight under-segmentation of the pancreas tumor remain.

%% file: table/main_table.tex
\begin{table*}[t]
\caption{Dice and normalized surface Dice (NSD) for colon and pancreas tumor. Bold indicates best performance. Significant improvements (2-tailed paired t-test, $p<0.05$) are denoted via $^*$. The promptable models use 1 point prompt per 3D volume.}
\label{main table}
\small
\begin{center}
    \begin{tabular}{ c c c c c c c c c  }
    \hline
    \hline
    %  \hline 3D UX-Net \cite{lee2023d}
     % & &  \vline & \multicolumn{6}{c}{\xxx{Methods}} \\
    % \multicolumn{2}{c}{Data \& metric} &  & nnU-Net & 3D UX-Net & Swin-UNETR & nnFormer & 3DSAM-adapter & Promise \\
    Dataset & Metric &  \vline & nnU-Net \cite{isensee2021nnu} & 3D UX-Net \cite{lee2023d} & nnFormer \cite{zhou2023nnformer} & Swin-UNETR \cite{tang2022self} & 3DSAM-adapter \cite{gong20233dsam} & ProMISe \\
    \hline
     
     % \multirow{2}{*}{{Colon}}
     % &Dice & \vline & 43.91 & 28.50 & 24.28 & 35.21 & 57.32 & \textbf{66.81}$^*$\\
     % &NSD  & \vline & 52.52 & 32.73 & 32.19 & 42.94 & 73.65 & \textbf{81.24}$^*$\\
     % \hline

     % \multirow{2}{*}{{Pancreas}}
     %  &Dice & \vline & 41.65 & 34.83 & 36.53 & 40.57 & 54.41 & \textbf{57.46}$^*$\\
     %  &NSD  & \vline & 62.54 & 52.56 & 53.97 & 60.05 & 77.88 & \textbf{79.76}$^*$\\
     % \hline

    %% below is my retrained, above is the number copied from cite{gong20233dsam}
     \multirow{2}{*}{{Colon}}
     &Dice & \vline & 45.60 & 23.07 &
     % 30.20 &
     21.36 & 37.23 & 57.32 & \textbf{66.81}$^*$\\
     &NSD  & \vline & 53.01 & 32.84 &
     % 42.25 &
     32.05 & 51.16 & 73.65 & \textbf{81.24}$^*$\\
     \hline

     \multirow{2}{*}{{Pancreas}}
      &Dice & \vline & 39.12 & 37.57 &
      % 34.03 &
      35.98 & 37.98 & 54.41 & \textbf{57.46}$^*$\\
      &NSD  & \vline & 57.66 & 55.25 &
      % 52.28 &
      53.45 & 56.42 & 77.88 & \textbf{79.76}$^*$\\
     \hline     
    % \multirow{2}{*}{{Colon}}
    %  &Dice & \vline & 45.60 & 23.07 & 30.20 & 21.36 & 37.23 & 57.32 & \textbf{67.23}$^*$\\
    %  &NSD  & \vline & 53.01 & 32.84 & 42.25 & 32.05 & 51.16 & 73.65 & \textbf{82.08}$^*$\\
    %  \hline
    %  \multirow{2}{*}{{Pancreas}}
    %   &Dice & \vline & 39.12 & 37.57 & 34.03 & 35.98 & 37.98 & 54.41 & \textbf{58.90}$^*$\\
    %   &NSD  & \vline & 57.66 & 55.25 & 52.28 & 53.45 & 56.42 & 77.88 & \textbf{80.43}$^*$\\
    %  \hline
     
  \end{tabular} 
\end{center}
\end{table*}

%% file: table/ablation_table.tex
\begin{table}[t]
\caption{Quantitative results of ablation study with single point prompt unless noted. R and C represent residual and concatenate fusions, and B indicates boundary loss. + shows the cumulative variants. Best viewed by individual sections.}
\label{ablation study}
\small
\begin{center}
    \begin{tabular}{  l| c c | c c}
    \hline
    \hline
    \multicolumn{1}{c}{} & \multicolumn{2}{c}{Colon}  & \multicolumn{2}{c}{Pancreas}\\
    \hline
     Method & Dice & NSD & Dice & NSD\\
     \hline
     
     baseline \cite{gong20233dsam} &  57.32 &  73.65   & 54.41  &  77.88\\

     + two adapters                &  61.61 &  73.88   & 56.08  &  77.89 \\
     + up-Conv                     &  62.92 &  77.62   & 55.37  &  77.38 \\
     \hline
     ProMISe-R                   &  63.67 &  79.96   & 55.15  &  79.02 \\
     
     ProMISe-R-B                 &  64.75 &  79.77   & 56.57  &  79.46 \\
     \hline
     ProMISe-C                   &  64.76 &  77.59   & 56.35  &  78.01 \\

     ProMISe-C-B (\textbf{proposed})                &  66.81 &  81.24   & 57.46  &  79.76 \\
     \hline

     baseline \cite{gong20233dsam} (10 prompts)               &  63.09 &  79.97   & 55.94  &  79.18 \\

     ProMISe-C-B (10 prompts)            &  67.28 &  81.63   & 58.05  &  80.36 \\
     \hline
     
  \end{tabular}
\end{center}
\end{table}

%% file: text/_5-conclusion.tex
% In this paper, we propose a promptable network, named ProMISe, designed for robust 3D tumor segmentation using pretrained weights from image foundation models. We evaluate on two public datasets, where our model consistently outperforms state-of-the-art methods across all tasks. Moreover, the critical role of the two adapters and boundary-aware loss techniques are demonstrated to achieve effective transfer learning between domains and improve the performance with ambiguous edges, respectively. Future work will aim to improve the efficiency through knowledge distillation. 
In this paper, we propose a promptable network, named ProMISe, designed for robust 3D tumor segmentation using pretrained weights from image foundation models. We evaluate on two public datasets, where our model consistently outperforms state-of-the-art methods across all tasks. Moreover, the critical role of the two adapters and boundary-aware loss techniques are demonstrated. Future work will aim to improve the efficiency through knowledge distillation. 

\noindent\textbf{Acknowledgments.}
This work was supported, in part, by NIH U01-NS106845, NSF grant 2220401.

%%% copied from other paper, maybe useful for ours journal

% The paper is structured as follows. Section II provides a
% review of related work, while Section III presents notations
% and preliminaries. In Section IV, we discuss our proposed kspace interpolation method and its corresponding theoretical
% guarantees. Implementation details are provided in Section
% V, followed by experimental results on several datasets in
% Section VI. We present our discussions in Section VII before
% concluding with some remarks in the final section VIII